\documentclass[aps,twocolumn,prd]{revtex4}
\tighten

\input epsf
\usepackage{graphicx}
\newcommand{\beq}{\begin{equation}}
\newcommand{\beqa}{\begin{eqnarray}}
		  \newcommand{\eeq}{\end{equation}}
\newcommand{\eeqa}{\end{eqnarray}}

\newcommand{\vect}[1]{\mbox{\boldmath${#1}$}}
\newcommand{\lmk}{\left(}
\newcommand{\rmk}{\right)}

\newcommand{\lkk}{\left[} 
\newcommand{\rkk}{\right]}

\newcommand{\ven}{\vect n}

\begin{document}

\title{ Estimating  detection rates of compact binary inspirals for 
networks of ground-based gravitational-wave detectors} 
%
%
\author{Naoki Seto}
\affiliation{Department of Physics, Kyoto University
Kyoto 606-8502, Japan
}
\date{\today}

%
%
%
%
\begin{abstract}

In a recent paper, Schutz proposed an analytical approximation for simplifying treatment of polarization angle and  conveniently 
evaluating  relative detection rates of compact binary 
inspirals for various networks of ground-based 
interferometers. 
We derived  relative event rates by strictly handling polarization 
angle and quantitatively examine validity of Schutz's approximation. 
The 
associated error of the approximation is rigorously shown to be less than 1.02\%, irrespective of 
details of the detector networks.

\end{abstract}
\maketitle

Currently, second-generation gravitational wave (GW) interferometers are being 
installed/constructed/planned around the world. Their most promising targets are 
inspirals of compact binaries, and various scientific prospects have been 
actively discussed.  

One of the primary measures for such studies is the detection rate of the 
binaries. While the overall rate is highly uncertain, due to limitation of 
our astronomical knowledge, the relative detection rates depend mainly on the 
geometry of the source-network configuration (see {\it e.g.} 
\cite{Schutz:2011tw,Cutler:1994ys}) for spatially homogeneous distribution 
 of sources. The relative rates play critical roles at 
examining performance of potential detector networks. The arguments related to 
the detection rates include dependence on  duty cycles of  constituent 
detectors, impacts of an  additional detector ({\it e.g.} LIGO-India), and
designing appropriate strategies ({\it e.g.} preferred survey directions) for 
counterpart searches with electromagnetic wave telescopes (see {\it e.g.} \cite{Schutz:2011tw,Raffai:2013yt,Searle:2006fp}).

However, the signal-to-noise ratios (SNRs) of individual binaries depend not 
only on their sky positions but also strongly on their orientations specified by 
 the inclination $I$ and polarization angle $\psi$ (explained below) \cite{Schutz:2011tw,Cutler:1994ys}. In order 
to make solid estimations of the relative rates, we have traditionally applied cumbersome methods such 
as Monte Carlo calculations for incorporating   
binary orientations.

For conveniently evaluating the relative event rates, Schutz recently proposed an analytical
approximation of taking certain average for the polarization angle $\psi$ 
\cite{Schutz:2011tw} (see {\it e.g.} \cite{Chen:2012qh} for its application). Then, only
 two dimensional integral with respect to the sky position is actually required 
 for the relative event rates. 
 But, in the paper,  the 
 accuracy of this approximation was left unexamined, with a comment that it can be tested by comparing with Monte 
 Carlo studies.

In this report, we analytically evaluate the relative rates with strictly 
handling the dependence on the polarization angle.  After deriving our final 
expression given in Eq.(8), we show how Schutz's 
approximation can be understood in our formulation and rigorously clarify its 
accuracy.

We assume coherent analysis of GWs with L-shaped interferometers 
labeled by $i=1,\cdots,m$ ($m$: total number of detectors). Due to the spin-2 
nature of GWs, we can generally 
express the responses of a detector $i$ to the incoming two polarization modes 
$+$  
and $\times$ as \cite{Schutz:2011tw,Cutler:1994ys}
\beqa
c_{i+}(\ven,\psi)&=&a_i(\ven)\cos2\psi+b_i(\ven)\sin2\psi,\\
c_{i\times}(\ven,\psi)&=&-a_i(\ven)\sin2\psi+b_i(\ven)\cos2\psi
\eeqa
with the polarization angle $\psi$ and  the source direction $\ven$.

For GW sources, we consider inspirals of  circular binaries that are assumed to 
have random positions and orientations, and  emit two 
polarization modes proportional to 
\beq
d_+(I)=\frac{I^2+1}2,~~d_\times(I)=I
\eeq
with  the inclination
$I\equiv \cos i$ ($i$: inclination angle). In Eqs.(1) and (2), the polarization 
angle $\psi$ fixes the azimuthal direction of the orbital angular momentum of  
binaries around the sky direction $\ven$.

Then, neglecting precession of orbital plane,  the coherent SNR  depends on the direction $\ven$ and orientation $(I,\psi)$ of a binary as
\beqa
SNR^2\propto \sum_{i=1}^m\lkk \lmk c_{i+}d_+    \rmk^2+\lmk c_{i\times}d_\times    
\rmk^2   \rkk \equiv f(\ven,I,\psi).
\eeqa
Here, applying trigonometric identities, the function $f$ can be expressed as
\beq
 f(\ven,\psi,I)=   \sigma(\ven) \lkk (d_+^2+d_\times^2) 
 +\epsilon(\ven) (d_+^2-d_\times^2)\cos4\psi'   \rkk \label{deff}
\eeq
with a shifted polarization angle $\psi'=\psi+\delta(\ven)$ and the two 
parameters $\sigma(\ven)$ and $\epsilon(\ven)$ that depend only on $\ven$ for a 
given detector network as
\beqa
\sigma(\ven) &\equiv& \sum_{i=1}^m\lkk 
a_i^2+b_i^2\rkk,\\
\epsilon(\ven)&=&\frac{\sqrt{\lkk \sum_{i=1}^m (a_i^2-b_i^2)\rkk^2+4(\sum_{i=1}^m a_ib_i)^2}}{\sigma(\ven)}.
\eeqa
The latter represents the asymmetry of the network sensitivities to the two 
polarization modes.
Using the Cauchy-Schwarz inequality, we can  show
$0\le \epsilon(\ven) \le 1$ with  the identity $\epsilon(\ven)=1$ for  a single 
detector network. Note that the expression (\ref{deff}) can be also found in \cite{Cutler:1994ys}.

For binaries with precessing orbital planes, the orientation angles $(I,\psi)$ 
change over time. Then, in Eq.(5), they should be regarded as appropriately 
averaged angles. This  mathematically complicate the problem. But our 
simple treatment above would be reasonable approximation at lease for double neutron stars \cite{Cutler:1994ys}.

Next, let us discuss the effective volume  detectable with the detector network 
by the coherent signal analysis.  
With respect to a fixed detection threshold for the coherent SNR, the maximum detectable distance 
$r_{max}$ scales as 
$
r_{max}\propto f(\ven,\psi,I)^{1/2}
$
for  given angular parameters $(\ven,\psi,I)$. Thus the effective volume 
associated with  a 
parameter space $d\ven d\psi dI$ is simply proportional to 
$
 f(\ven,\psi,I)^{3/2} d\ven d\psi dI $.

By integrating out the source orientation angles $(\psi,I)$, the effective  
volume (equivalently relative detection rate) for a 
given solid angle $d\ven$ is proportional to
\beq
\sigma(\ven)^{3/2}g(\epsilon(\ven))d\ven,
\eeq
where the new function $g(\epsilon)$ is defined by 
\beqa
g(\epsilon)\equiv&& \frac1{2^{5/2}\pi}\int_0^{\pi}d\nonumber\psi\int_{-1}^1dI \big[ (d_+^2+d_\times^2) \\
& & \ \ \ \ +\epsilon (d_+^2-d_\times^2)\cos4\psi   \big]^{3/2}\label{defg}
\eeqa
with the  normalization factor $2^{5/2}\pi$ given for the double integrals with 
$d_+(1)=d_\times(1)=1$ (corresponding to face-on binaries).

The function $g(\epsilon)$ monotonically increases  in the 
relevant range $0\le \epsilon \le 1$ with
\beq
g(0)=0.290451,~~g(1)=0.293401=1.010125\times g(0).  \label{r1}
\eeq
The numerical value $g(0)$ is identical to that given in \cite{Schutz:2011tw}. 
 By perturvatively expanding Eq.(9),  we also have
\beq
g_{exp}(\epsilon)=0.290451 (1+0.00978\epsilon^2+0.00026\epsilon^4+O(\epsilon^6)) \label{r2}
\eeq
with accuracy of $\left|{g_{exp}(\epsilon)}/{g(\epsilon)}-1\right|<10^{-4}$ 
(dropping $o(e^4)$ terms) in  the range $0\le \epsilon \le 1$.
We can anticipate the observed weak dependence on $\epsilon$, considering that (i) the 
integral (\ref{defg}) becomes constant at the power index 1 close the 
original one 3/2, and (ii) we have  $g'(0)=0$ due to the symmetry of 
the integrand.

Now we discuss Schutz's approximation. In our formulation,  it corresponds to taking $\psi$ average at the stage of 
Eq.(5) before the nonlinear operation $[\cdots]^{3/2}$ in Eq.(9).  This is 
actually equivalent to putting $\epsilon(\ven)=0$ in Eq.(9) and the resultant 
expression is identical to 
\beq
\sigma(\ven)^{3/2}g(0)d\ven
\eeq
 in contrast to Eq.(8) obtained in our strict derivation. 

But our results (\ref{r1}) and (\ref{r2}) show that,  for evaluating the 
relative  detection rates,   disregard of the 
$\epsilon$ dependence (thus only with the leading term in Eq.(11)) is an excellent 
approximation with error less than 
1.02\%. Since the integrands in Eqs.(8) and (12) are non negative, the quoted 
accuracy is also valid for the final results after the sky average.
If necessary, we can readily include the $\epsilon$-dependence (\ref{r2}) for $g(\epsilon)$.

Within the guaranteed accuracy of $1.02\%$, we can now justify evaluating the relative 
 detection rate in the solid angle $d\ven$ simply by
\beq
\sigma(\ven)^{3/2} d\ven
\eeq
or the total rate by
\beq
\int_{4\pi}  \sigma(\ven)^{3/2} d\ven
\eeq
without resorting to cumbersome Monte-Carlo calculations to handle the orientations
of the binaries. 

If the detectors $i=1,\cdots,m$ have different 
sensitivities (or equivalently horizon distances), we can straightforwardly 
apply our results by introducing appropriate weights to the response functions $(a_i,b_i)$. 
Furthermore, the form (5) can be derived even in the presence of certain correlated noises 
between detectors   with the corresponding functions $\sigma(\ven)$ and 
$\epsilon(\ven)$  \cite{Cutler:1994ys}, and our results are unchanged also in 
such cases.
 
This work was supported by JSPS (24540269) and MEXT (24103006).



\end{document}